# High-Responsivity Photodetection by Self-Catalyzed Phase-Pure P-GaAs Nanowire

Hassan Ali[1,5,†], Yunyan Zhang[2,†], Jing Tang[1], Kai Peng[1], Sibai Sun[1], Yue Sun[1], Feilong Song[1], Attia Falak[3,4], Shiyao Wu[1], Chenjiang Qian[1], Meng Wang[1], Zhanchun Zuo[1], Kui-Juan Jin[1,5], Ana M. Sanchez[6], Huiyun Liu[2] and Xiulai Xu[1,5,7*]

[1]*Beijing National Laboratory for Condensed Matter Physics, Institute of Physics, Chinese Academy of Sciences, Beijing 100190, China*
[2]*Department of Electronics and Electrical Engineering, University College London, London WC1E 7JE, United Kingdom*
[3]*National Centre for Nanoscience and Technology, Chinese Academy of Sciences, Beijing 100190, China*
[4]*Department of Physics, University of the Punjab, Quaid-e-Azam Campus, Lahore, 54000, Pakistan*
[5]*School of Physical Sciences, University of Chinese Academy of Sciences, Beijing 100190, China*
[6]*Department of Physics, University of Warwick, Coventry CV4 7AL, United Kingdom*
[7]*CAS Center for Excellence in Topological Quantum Computation, University of Chinese Academy of Sciences, Beijing 100190, China*

[†] These Authors contribute equally to this work
* Author to whom correspondence should be addressed. Email for correspondence: xlxu@iphy.ac.cn

Defects are detrimental for optoelectronics devices, such as stacking faults can form carrier-transportation barriers, and foreign impurities (Au) with deep-energy levels can form carrier traps and non-radiative recombination centers. Here, we first developed self-catalyzed p-type GaAs nanowires (NWs) with pure zinc blende (ZB) structure, and then fabricated photodetector made by these NWs. Due to absence of stacking faults and suppression of large amount of defects with deep energy levels, the photodetector exhibits room-temperature high photo responsivity of $1.45 \times 10^5$ A W$^{-1}$ and excellent specific detectivity (D*) up to $1.48 \times 10^{14}$ Jones for low-intensity light signal of wavelength 632.8 nm, which outperforms previously reported NW-based photodetectors. These results demonstrate that these self-catalyzed pure-ZB GaAs NWs to be promising candidates for optoelectronics applications.

**Keywords:** GaAs nanowires, self-catalyzed growth, phase-pure crystal structure, photodetector, photo-responsivity.

In recent years, semiconductor nanowires (NWs) have received tremendous attention because of their enormous potential applications in nanoscale electronics and optoelectronics devices such as photodetectors, single electron transistors, single photon detectors, tunneling diodes and nano lasers.[1-8] Reduced dimensionality, large surface to volume ratio and strong light matter interaction make the NWs a solid candidate for optoelectronic applications.[2, 9, 10] Photodetector is one of the optoelectronic devices which has large range of applications in environmental monitoring, industrial quality control and optical communications.[11-13] Furthermore, high photo responsivity, signal-to-noise ratio, specific detectivity and fast response time are the desired parameters for photodetectors.[11, 12]

Photodetectors fabricated using semiconductor NWs have been a source of motivation for researchers due to their high quantum efficiency and responsivity.[14] Among all semiconducting NWs, III-V semiconductor NWs are promising candidates for photodetectors because of their high absorption coefficient and wide tunable bandgaps.[3] Photodetectors based on III-V semiconductor NWs have been fabricated in different configurations such as core-shell nanostructures[1, 15], alloys[16-18] and hetero structures[19, 20]. Among all III-V NWs, gallium arsenide (GaAs) NWs have gained immense attention for detection application over recent years because of their high light-to-electricity conversion efficiency, moderate direct bandgap (1.42 eV) and high compatibility with Si technology, which in turn make them suitable for various outstanding optoelectronic applications like solar cells, photodetectors, p-n diodes and field effect transistors.[19, 21-30]

The performance of photodetectors depends largely on the growth techniques of NWs. To the best of our knowledge, most of photodetectors are fabricated based on NWs which are synthesized by foreign catalyst assisted techniques which compromises photodetection parameters.[17, 23, 31-33] The foreign catalyzed metal (e.g. Au) significantly degrades the optoelectronic properties of NWs by incorporating foreign catalyst impurities into NWs during growth process which generate deep trap levels behaving as efficient recombination centers.[7, 34] These NWs are also not well-matched with the complementary metal-oxide-semiconductor (CMOS) engineering standards because of the inclusion of defect states.[35] Therefore, self-induced growth (self-catalyzed growth) techniques to synthesize the NWs have become intensively vital because they outmaneuver unintentional incorporation of foreign impurities which enhances the optoelectronic features of NWs.[7, 18, 34-36]

However, self-catalyzed NWs are commonly observed to have stacking faults, intermixing of zinc blende (ZB) and wurtzite (WZ) structures, which are difficult to be eliminated due to small growth window. This intermixing happens when NW is grown in ⟨1 1 1⟩ crystal direction of cubic cell or ⟨0 0 0 1⟩ direction of hexagonal cell, because of small nucleation energy difference between them (smaller than 25 meV per ortho pair).[37-39] They can significantly degrade the device performance, such as acting as scattering centers for carriers and form carrier-transportation barriers.[40-42] It has been observed that the increase in stacking fault density decreases carrier mobility of GaAs NWs from ~2250 to 1200 cm$^2$ V$^{-1}$ s$^{-1}$.[43] Moreover, Wallentin *et al* reported that the ZB segments in WZ InP NWs can act as traps for carriers.[44] For un-doped InP NWs, the trapped carrier concentration can be as high as $4.6 \times 10^{18}$ cm$^{-3}$, which leads to low conductivities and motilities. Therefore, elimination of stacking faults is of great importance for improving the carrier transportation and hence the device performance.

In this work, we report on self-catalyzed phase-pure single GaAs NW based photodetector (PD) and its photodetection characteristics. At room temperature, photodetector exhibits a high photo responsivity (R) of 1.45 x 10$^5$ A W$^{-1}$, photoconductive gain (G) of 2.85 x 10$^7$ % and significant spectral detectivity (D*) up to 1.48 x 10$^{14}$ Jones for light of wavelength 632.8 nm with weak intensity of 0.03 mW/cm$^2$, which outperforms previously reported photodetectors based on NWs[16-18, 23] without ferroelectric polymer layer.[14]

The un-doped and beryllium-doped GaAs NWs were grown by self-catalyzed mode.[35, 45] Both of them have a diameter of 50-60 nm and are uniform along the length (Figure 1 (a) and (b)). However, there are differences in crystal quality. The beryllium-doped NWs have pure-ZB crystal structure up to the top most bilayer (Figure 1 (c-f)), while the un-doped NWs have high density of planar defects along the length (Figure 1(g-j)). The suppression of WZ nucleation during growth can be explained in term of the lowering of droplet supersaturation and/or surface energy by the formation of Be-Ga alloy droplets.[46-48] As can be seen in Figure 1 (b) and (g), Ga droplets from un-doped NWs are round and at the exact top of the tip, while those from Be-doped NWs are displaced from the nanowire center (Figure 1(a) and (c)), indicating the change of vapor-liquid and liquid-solid interface energies and hence confirming the formation of Be-Ga alloy droplets by doping.

The photoluminescence spectra of p-doped GaAs NWs reveal only single peak emission with peak intensity located at wavelength around 833 nm (1.49 eV) which is due to the conduction band to acceptor recombination (green line in Figure 2 (a)).[49, 50] The asymmetric line shape indicates that distribution of free carriers can be described by Boltzmann statistics because of the minimal inhomogeneity at the band edge energy.[51] There is no other defect related emission at longer wavelength. This is in stark contrast to un-doped GaAs NWs which have very weak band-to-band emission but very strong emission at longer wavelength, indicating the existence of a large amount of defects with deep energy levels (blue line in Figure 2(a)). These phenomena can be explained by doping-induced change in band bending. The NWs have a large surface-to-volume ratio and strongly influenced by high-density surface states. For un-doped GaAs NWs, surface states can act as traps for electrons and fully deplete the NWs, leading to the upward bending of conduction band (CB) and valence band (VB) as illustrated in left part of Figure 2 (b).[52, 53] The photon generated holes are diffused efficiently and accumulated at the NW surface because the surface is a low-energy well for holes. On the other hand, electrons move to surface by tunneling through and climbing over the barrier due to high mobility and small effective mass. As a result, photon generated electrons and holes are recombined at the surface, leading to weak band-to-band emission but showing emission at longer wavelength. For p-type NWs, the surface states are donor-like and can trap holes, leading to a negatively charged surface depletion region and causing downward bending of CB and VB bands as illustrated in right part of Figure 2 (b).[54] This can effectively confine the photo-generated holes in the NW center, because holes with large effective mass and low mobility are difficult to tunnel through or climb over the barrier like electrons. As a result, doped NWs have larger band-to-acceptor emission and lower surface-state emission. Therefore, p-doping can reduce the influence of NW surface states and carrier loss.

Due to the advantages of p-type doping mentioned above, photodetector has been fabricated using p-GaAs single NW. The SEM image of single NW PD has been shown in the inset of Figure 3 (a). Prior to the photodetection measurement, the electrical transport characteristics of the device were studied in field effect transistor configuration (schematic is shown in the inset of Figure 3 (b)) to confirm the nature of nanowire and to explore the contact behavior between nanowire and metal electrode. Figure 3 (a) shows the output characteristics curves of GaAs NW FET as a function of back gate bias at room temperature. The gate bias Vg has been varied from -10 V to 10 V with interval of 5 V. The non-linear behavior of output characteristics depicts the presence of schottky junction between metal contacts and NW. The slight non-symmetric trend of output curves shows the presence of schottky junction with minor difference in barrier height between GaAs NW and metal electrodes at both ends of NW. The schottky junction between metal electrodes and NW is advantageous to achieve better photo-detection characteristics as compared to ohmic contacts.[4] The output characteristics curves of NW FET confirm p-type semiconductor properties in which drain current decreases (increases) with an increase in positive (negative) gate bias. Under negative gate bias, the accumulation of holes into NW channel increases, which thus increases the drain current. In case of positive gate bias, the accumulation of electrons occurs in channel which reduces the drain current. The drain current dependence on back gate voltage (transfer characteristics curves) for different drain voltages, at room temperature has also been shown in Figure 3 (b). The transfer characteristics curves show the decrease in drain current with back gate bias which also confirms the p-type conductivity of NW. The calculated field effect mobility of GaAs NW FET is 0.05 cm$^2$/V-sec and corresponding carrier density is 3.36 x 10$^{17}$/cm$^3$. The detailed calculations for mobility and concentration are given in supporting information S1. The small diameter (~50nm) makes the formation of Ohmic contact difficult. Due to the influence of schottky junction between NW and metal electrodes, the calculated field effect mobility of the p-GaAs NW is quite low. Furthermore field effect mobility largely depends on the geometry of device and surface states density.

The room-temperature photo response performance of GaAs NW FET photodetector has been investigated in terms of its output characteristics for different light intensities. Figure 4 (a) represents the output characteristics of GaAs NW photodetector in dark and under light illumination at zero gate bias. The dark current at zero gate bias under forward and reverse biased is 8 nA and 4 nA, respectively. The difference is caused by different Schottky barrier heights at two ends of NW. The photo-current increases with light illumination because of the increased density of photo generated carriers. The charge carrier′s photo generation efficiency is proportional to the absorbed photon flux. These carriers contribute in conductivity, results an increase in drain current. The net photocurrent is defined by the expression, $I_{ph} = I_{light} - I_{dark}$, where $I_{dark}$ is the current before illumination and $I_{light}$ is the current under illumination. The value of photocurrent obtained at $V_g = 0$ V and $V_{sd} = -5$ V is 15 nA for light intensity of 87.9 mW/cm$^2$. The inset of Figure 4 (a) shows the photocurrent of NW device on linear scale for different light intensities.

Figure 4 (b) shows the dependence of photocurrent on light intensity at zero gate voltage for different drain bias and it follows power law, $I_p \alpha P^\gamma$ where exponent γ describes the response of photocurrent with light intensity. The low intensity light would be absorbed in depletion region creating electron hole pairs. These generated electron hole pairs then move under built-in electric field increasing the photocurrent linearly. At low light intensity, the strong built-in-electric field around schottky junction can effectively separate the photo-generated electron-hole pairs and hence can suppresses their recombination, leading to high photon response. Under strong illumination, the newly-balanced state makes the carrier concertation at the schottky junction higher leading to the reduction of the depletion width. Therefore, the reduced built-in-electric field weakens the carrier separation efficiency, resulted in lower photon response. Therefore, with further increase in light intensity, no further increase in photocurrent is observed and saturation is achieved.[4, 55] The more clear explanation of this mechanism has been illustrated later with help of band structure of metal-semiconductor junction. The values of exponent γ have been extracted by non-linear curve fitting for different drain voltages and it yields γ= 0.34, 0.43, 0.59 and 0.73 for $V_{sd}$ of 1.2 V, 2.4 V, 3.6 V and 4.8 V respectively. The non-unity values of γ suggest the complex processes such as electron-hole pair generation, recombination and trapping of charge carriers in GaAs NWs.[2, 56]

Figure 4 (c) represents the ratio of photocurrent to dark current i.e. photo response ratio ($I_{ph}/I_{dark}$) of photodetector against drain bias at zero $V_g$. The photo response ratio increases with increase in light intensity because of the contribution of more photo generated carriers at higher light intensities. The ratio $I_{ph}/I_{dark}$ is approximately 4.5 at drain bias of -5 V for light intensity of 87.9 mW/cm$^2$. The low value of photo response ratio might be attributed to large dark current but the value in our case is better than previously reported NW photodetectors due to the absence of stacking faults and the suppression of a large amount of defects with deep energy levels.[14, 33] The time dependence photo response of GaAs single NW photodetector has been measured by periodically switching on and off 632.8 nm light with intensity of 87.9 mW/cm$^2$ at $V_{ds}$ = 5 V and $V_g$ = +15 V. As depicted from Figure 4 (d), the device current rises up abruptly and reaches a steady state of 19 nA under incident light and then drops down to its initial current value of 9 nA in absence of illumination. The positive gate voltage initially depletes the carriers in nanowire resulting a decrease in drain current, later the concentration becomes constant with time, and stable and reproducible characteristics have been achieved in GaAs NW photodetector.

Spectral responsivity (R), defined as the photocurrent generated when unit power of light intensity shines on effective area of NW and can be expressed as,[2, 14]

$$R = I_{ph}/PA \qquad (1)$$

Where *P* is the incident light intensity and *A* is the effective area of NW. The effective area of the laser spot was measured to follow Gaussian distribution and the width of peak position of distribution is about 10 microns, therefore we considered the uniform light shine on the NW. The optical gain (G) is the vital parameter which is related to the number of electron-hole pairs excited by absorption of one photon. It also explains the efficiency of the carrier transport. The optical gain can be expressed by the following equation,[2, 14]

$$G = R*hc/\lambda e \qquad (2)$$

Where *h* is the plank constant, *e* is the electronic charge, *c* is the speed of light and *λ* is the wavelength of irradiated light. Another figure-of-merit of photodetector is its detectivity (*D*\*) which portrays the proficiency to the smallest detectable signal and can be defined as,[3, 14]

$$D^* = (A\Delta f)^{1/2}/(NEP) \qquad (3)$$

Where *Δf* is the electrical bandwidth in Hz and *NEP* is the noise equivalent power. In case of small noise from dark current, the specific detectivity can also be written as

$$D^* = R\,(A/2eI_{dark})^{1/2} \qquad (4)$$

Where $I_{dark}$ is the current before illumination. For incident light intensity of 0.03 mW/cm² at $V_{sd}$ equals -5 V, the value of spectral responsivity has been found to be 1.45 x 10⁵ A/W. The high value of photo responsivity is attributed to the strong built-in electric field at GaAs NW/metal electrode schottky junction. The corresponding values of optical gain and spectral detectivity are 2.85 x 10⁷ % and 1.48 x 10¹⁴ Jones respectively which are superior to the previous NW detectors with bare NW surface.[16-19] The outstanding performance is due to: (1) the absence of foreign metal contamination that can reduce the carrier traps and non-radiative recombination centers inside NWs; (2) downward band bending that can reduce the influence of surface state and hence the loss of carriers; (3) single-phase crystal quality of NWs that reduces scattering centers, and hence more efficient carrier collection at the contact region.

Field effect mobility and photodetection parameters largely depend on the device geometry and configuration. In contrast with the mobility, the photodetection parameters such as responsivity and detectivity have large values in our case. The device geometry is optimized in such a way as to achieve good photodetection parameters. The channel length is kept small to reduce the electron-hole recombination and to gain high responsivity with compromised value of the mobility. The other reason of low mobility, "the schottky barrier between metal electrode and nanowire" is responsible for high responsivity which separates photo-generated electron-hole pairs and hence results in high responsivity.

The spectral responsivity R as a function of incident light intensity for different drain bias has been shown in Figure 5 (a). It is revealed from figure that spectral responsivity decreases with increasing light intensity for both positive and negative bias voltages. It can be explained as follows: when light irradiates on semiconducting NW, three different processes namely, electron-hole pair generation, electron-hole pair recombination and motion of generated carriers under strong built-in electric field take place.[2, 14, 33] Photo generated carriers are proportional to light intensity and its absorption into the material. At low light intensity, strong built-in electric field reduces the rate of recombination resulting in high spectral responsivity. However at higher light intensities, the increased rate of recombination because of reduced built-in-potential lowers the responsivity.[14, 33, 56] It is also obvious from the Figure 5 (a) that the photo responsivity is higher for $V_{sd}$ = -5 V than for $V_{sd}$ = 5 V. It is because of different schottky barrier heights between metal and GaAs NW during forward and reverse bias. The behavior of schottky junction between GaAs NW and metal electrode under illumination of light, and under different biasing conditions has also been elaborated with the help of band diagram as shown in Figure 6. The work function of metal contact chromium ($\varphi_{cr}$) is 4.5 eV which is smaller than the work function of p-type GaAs $\varphi_{GaAs}$ (5.1 eV). Upon contact between metal electrode and GaAs NW, the negative work function difference ($\varphi_{cr} - \varphi_{GaAs}$) results in a downward band-bending at the interface under thermal equilibrium as shown in Figure 6 (a). It thus builds a layer of negative charges in semiconductor region and positive charges in metal electrode near the interface, results in the formation of schottky barrier at the interface.[4, 57-59] The device current through schottky contacts is extremely sensitive to the schottky barrier height and its width.[60] The height of schottky barrier in p-type semiconductors is defined as $\varphi_B = E_g - (\varphi_{cr} - \chi)$ where $\chi$ is the electron affinity of GaAs NW and its value is 4.07 eV.[5, 20, 61]

When the junction is reverse biased as shown in Figure 6 (b), the movement of holes from semiconductor to metal electrode decreases, also the applied field and built-in electric field are parallel to each other. These factors contribute in increasing the barrier width and strengthening the built-in electric field which separates the photo generated electron hole pairs resulting in high photocurrent and photo responsivity. Under forward bias, holes, the dominant carriers in p-type GaAs NWs tend to transport from GaAs NWs to the metal electrode reducing the schottky barrier (not shown). Under light illumination and forward biasing of schottky junction, the applied electric field is antiparallel to the built-in electric field of schottky barrier, which lowers the built-in electric field, increases electron hole recombination process which in turns reduces photocurrent and photo responsivity.

The inset of Figure 5 (a) shows the variation of spectral responsivity of GaAs NW PD with applied bias at light intensity of 0.03mW/cm$^2$. The spectral responsivity increases with increase in bias voltage. Specific detectivity of PD with incident light intensity has been shown Figure 5 (b) for different bias voltages. The detectivity also decreases as light intensity increases but at low intensities, we achieved high value of detectivity. A comprehensive comparison of these critical parameters of photodetectors using single III-V NWs has been performed and shown in Table 1. The high parameters achieved in our GaAs single NW photodetector make it suitable for detecting weak signals which has broad applications.[14]

**Table 1:** Comparison of responsivity, optical gain and detectivity of GaAs NW PD with previous reported work.

| Material | Dark Current | Responsivity (A/W) | Optical Gain (%) | Detectivity (Jones) | Reference |
|---|---|---|---|---|---|
| InAs NW | ~ 5 µA at 10 V | 8.4 x 10$^4$ | 1.96 x 10$^6$ | | [2] |
| GaAsSb NW | ~ 200 nA at 1V | 1.7 x 10$^3$ | 1.62 x 10$^5$ | | [16] |
| GaAsSb NW | ~ 320nA at 0.5V | 2.37 | | 1.08 x 10$^9$ | [17] |
| InAs NW | ~-2nA at -2 V | 5.3 x 10$^3$ | | | [4] |
| GaAsSb NW | ~ 90 nA at -3V | 1.5 x 10$^3$ | | | [18] |
| GaAs NW | ~ 80 nA at -5V | | 2x 10$^6$ | | [23] |
| InAs NW | | 40 | | 2 x 10$^{12}$ | [62] |
| GaAs NW | ~-4nA at -5 V | 1.45 x 10$^5$ | 2.85 x 10$^7$ | 1.48 x 10$^{14}$ | Our work |

In conclusion, we have successfully grown self-catalyzed p-type GaAs NWs with pure ZB structure, which has been confirmed with high-resolution TEM and photoluminescence spectroscopy. By using these phase-pure p-type GaAs NWs, we have demonstrated GaAs single NW based photodetector which has extraordinary characteristics compared with other conventional foreign-catalyzed assisted grown NWs based PDs. GaAs NW based PD reveals ultrahigh photo responsivity of 1.45 x 10$^5$ A W$^{-1}$, remarkable photoconductive gain of 2.85 x 10$^7$ % and exceptional spectral detectivity (D*) up to 1.48 x 10$^{14}$ Jones at 632.8 nm with low excitation laser intensity of 0.03 mW/cm$^2$. These outstanding features of PD based on self-catalyzed grown GaAs NW make the NWs striking constituent for plenteous electronics and optoelectronics devices.

**Experimental Section:**

*NW growth:* The self-catalyzed GaAs NWs were grown directly on p-type Si (111) substrates by means of solid-source III−V molecular beam epitaxy (MBE). GaAs NWs were grown with a Ga beam equivalent pressure, V/III flux ratio, substrate temperature, growth duration and nominal doping concentration (characterized in thin film growth) of $8.41\times10^{-8}$ Torr, 50, ~630°C, 1 hour and $1.6$~$6.4\times10^{18}$/cm$^3$, respectively. The substrate temperature was measured by pyrometer.

*Photoluminescence measurement:* The photoluminescence spectrum of single NWs has been measured using conventional confocal micro-PL system under vacuum at 20 K with a 532 nm green laser with a spot size having a diameter of about 1-2 μm.

*Detector fabrication:* For the fabrication of GaAs single NW based FET, an n-type Si wafer coated with thermally grown 300 nm SiO$_2$ has been used as a substrate. As grown GaAs NWs were suspended in isopropyl alcohol (IPA), sonicated and drop casted on SiO$_2$/Si substrate. After mapping single GaAs NW by SEM, electrical contacts on both ends of NW were patterned by using electron beam lithography. Contact metals electrodes (Cr/Au 20/80nm) were deposited using thermal evaporation followed by lift off process in hot acetone. The highly-doped Si substrate has been used as a back-gate electrode.

*Characterization:* The electrical performance of individual NW FET device has been characterized at room temperature using home-made dipstick connected with semiconductor device parameter analyzer. The measurements for GaAs single NW based photodetector were also carried out at room temperature using semiconductor characterization system. He-Ne laser having wavelength of 632.8 nm was used with different intensities ranging from 0.03 mW/cm$^2$ to 87.9 mW/cm$^2$ to observe the photo response of detector.

*Focus ion beam/scanning electron microscope (FIB/SEM):* The NW morphology was measured with a Zeiss XB 1540 FIB/SEM system.

*Transmission electron microscopy (TEM):* Simple scraping of the NWs onto a lacey carbon support was used to prepare TEM specimens. The TEM measurements were performed on JEOL 2100 and doubly−corrected ARM200F microscopes, both operating at 200 kV.

**Acknowledgements:**

This work was supported by the National Basic Research Program of China under Grant No. 2014CB921003; the National Natural Science Foundation of China under Grant No. 11721404, 51761145104 and 61675228; the Strategic Priority Research Program of the Chinese Academy of Sciences under Grant No. XDB07030200, and XDPB0803, and the CAS Interdisciplinary Innovation Team; the Leverhulme Trust and EPSRC research grant (EP/P000886/1). HA is supported by CAS-TWAS President's Fellowship Program.

**References:**

**References:**

[1]   N. Erhard, S. Zenger, S. Morkötter, D. Rudolph, M. Weiss, H.J. Krenner, H. Karl, G. Abstreiter, J.J. Finley, G. Koblmüller, *Nano Letters* **2015** *15*, 6869.
[2]   C. H. Kuo, J. M. Wu, S. J. Lin, W. C. Chang, *Nanoscale Research Letters* **2013**, *8*, 327.
[3]   R. LaPierre, M. Robson, K. Azizur-Rahman, P. Kuyanov, *Journal of Physics D: Applied Physics* **2017**,*50*, 123001.
[4]   J. Miao, W. Hu, N. Guo, Z. Lu, X. Zou, L. Liao, S. Shi, P. Chen, Z. Fan, J.C. Ho, *ACS Nano* **2014**, *8*, 3628.
[5]   M. Heiss, S. Conesa-Boj, J. Ren, H. H. Tseng, A. Gali, A. Rudolph, E. Uccelli, F. Peiró, J.R. Morante, D. Schuh, *Physical Review B* **2011**, *83*, 045303.
[6]   Y.A. Du, S. Sakong, P. Kratzer, *Physical Review B* **2013**, *87*, 075308.
[7]   J.V. Holm, H.I. Jørgensen, P. Krogstrup, J. Nygård, H. Liu, M. Aagesen, *Nature Communications* **2013**, *4*, 1498.
[8]   X. Xu, H. Baker, D.A. Williams, *Nano Letters* **2010**, *10*, 1364.
[9]   J.O. Island, S.I. Blanter, M. Buscema, H.S. van der Zant, A. Castellanos-Gomez, *Nano Letters* **2015**, *15*, 7853.
[10]  M.W. Chen, C. Y. Chen, D. H. Lien, Y. Ding, J. H. He, *Optics Express* **2010**, *18*, 14836.
[11]  J. Zhou, Y. Gu, Y. Hu, W. Mai, P. H. Yeh, G. Bao, A.K. Sood, D.L. Polla, Z.L. Wang, Applied Physics Letters **2009**, *94*, 191103.
[12]  V. Logeeswaran, J. Oh, A.P. Nayak, A.M. Katzenmeyer, K.H. Gilchrist, S. Grego, N.P. Kobayashi, S. Y. Wang, A.A. Talin, N.K. Dhar, *IEEE Journal of Selected Topics in Quantum Electronics* **2011**, *17*, 1002.
[13]  M.H. Huang, S. Mao, H. Feick, H. Yan, Y. Wu, H. Kind, E. Weber, R. Russo, P. Yang, *Science* **200**1, *292*, 1897.
[14]  D. Zheng, J. Wang, W. Hu, L. Liao, H. Fang, N. Guo, P. Wang, F. Gong, X. Wang, Z. Fan, *Nano Letters* **2016**, *16*, 2548.
[15]  E.M. Gallo, G. Chen, M. Currie, T. McGuckin, P. Prete, N. Lovergine, B. Nabet, J.E. Spanier, *Applied Physics Letters* **2011**, *98*, 241113.

[16]  L. Ma, X. Zhang, H. Li, H. Tan, Y. Yang, Y. Xu, W. Hu, X. Zhu, X. Zhuang, A. Pan, *Semiconductor Science and Technology* **2015**, *30*, 105033.
[17]  Z. Li, X. Yuan, L. Fu, K. Peng, F. Wang, X. Fu, P. Caroff, T.P. White, H.H. Tan, C. Jagadish, *Nanotechnology* **2015**, *26*, 445202.
[18]  J. Huh, H. Yun, D.-C. Kim, A.M. Munshi, D.L. Dheeraj, H. Kauko, A.T. van Helvoort, S. Lee, B.O. Fimland, H. Weman, *Nano Letters* **2015**, *15*, 3709.
[19]  A.M. Munshi, D.L. Dheeraj, V.T. Fauske, D.-C. Kim, A.T. van Helvoort, B.-O. Fimland, H. Weman, *Nano Letters* **2012**, *12*, 4570.
[20]  Z. Xu, S. Lin, X. Li, S. Zhang, Z. Wu, W. Xu, Y. Lu, S. Xu, *Nano Energy* **2016**, *23*, 89.
[21]  H. Xia, Z. Y. Lu, T. X. Li, P. Parkinson, Z. M. Liao, F. H. Liu, W. Lu, W. D. Hu, P. P. Chen, H. Y. Xu, , *ACS Nano* **2012**, *6*, 6005.
[22]  M.A. Seyedi, M. Yao, J. O'Brien, S. Wang, P.D. Dapkus, *Applied Physics Letters* **2013**, *103*, 251109.
[23]  H. Wang, *Applied Physics Letters* **2013**, *103*, 093101.
[24]  N. Han, Z. X. Yang, F. Wang, G. Dong, S. Yip, X. Liang, T.F. Hung, Y. Chen, J.C. Ho, *ACS Applied Materials & Interfaces* **2015**, *7*, 20454.
[25]  H. Shu, X. Yang, P. Liang, D. Cao, X. Chen, *The Journal of Physical Chemistry C* **2016**, *120*, 22088.
[26]  M. Orrù, V. Piazza, S. Rubini, S. Roddaro, *Physical Review Applied* **2015**, *4*, 044010.
[27]  M. Mikulics, M. Marso, S. Mantl, H. Lüth, P. Kordoš, *Applied Physics Letters* **2006**, *89*, 091103.
[28]  J. Yoon, S. Jo, I.S. Chun, I. Jung, H.S. Kim, M. Meitl, E. Menard, X. Li, J.J. Coleman, U. Paik, *Nature* **2010**, *465*, 329.
[29]  J.A. Czaban, D.A. Thompson, R.R. LaPierre, *Nano Letters* **2008**, *9*, 148.
[30]  P.K. Mohseni, G. Lawson, C. Couteau, G. Weihs, A. Adronov, R.R. LaPierre, *Nano Letters* **2008**, *8*, 4075.
[31]  B.S. Sørensen, M. Aagesen, C.B. Sørensen, P.E. Lindelof, K.L. Martinez, J. Nygård, *Applied Physics Letters* **2008**, *92*, 012119.
[32]  N. Han, J.J. Hou, F. Wang, S. Yip, Y.T. Yen, Z.X. Yang, G. Dong, T. Hung, Y. L. Chueh, J.C. Ho, *ACS Nano* **2013**, *7*, 9138.
[33]  L. Zhang, X. Geng, G. Zha, J. Xu, S. Wei, B. Ma, Z. Chen, X. Shang, H. Ni, Z. Niu, *Materials Science in Semiconductor Processing* **2016**, *52*, 68.
[34]  D. Rudolph, S. Hertenberger, S. Bolte, W. Paosangthong, D.e. Spirkoska, M. Döblinger, M. Bichler, J.J. Finley, G. Abstreiter, G. Koblmüller, *Nano Letters* **2011**, *11*, 3848.
[35]  Y. Zhang, A.M. Sanchez, Y. Sun, J. Wu, M. Aagesen, S. Huo, D. Kim, P. Jurczak, X. Xu, H. Liu, *Nano Letters* **2016**, *16*, 1237.
[36]  P. Krogstrup, R. Popovitz-Biro, E. Johnson, M.H. Madsen, J. Nygård, H. Shtrikman, *Nano Letters* **2010**, *10*, 4475.
[37]  T. Akiyama, K. Sano, K. Nakamura, T. Ito, *Japanese Journal of Applied Physics* **2006**, *45*, L275.
[38]  M. Bukała, M. Galicka, R. Buczko, P. Kacman, H. Shtrikman, R. Popovitz-Biro, A. Kretinin, M. Heiblum, *AIP Conference Proceedings* **2010**, *1199*, 349.
[39]  T. Ito, *Japanese Journal of Applied Physics* **1998**, *37*, L1217.
[40]  M. Schroer, J. Petta, *Nano Letters* **2010**, *10*, 1618.
[41]  C. Thelander, P. Caroff, S. Plissard, A.W. Dey, K.A. Dick, *Nano Letters* **2011**, *11*, 2424.
[42]  M.J. Sourribes, I. Isakov, M. Panfilova, H. Liu, P.A. Warburton, *Nano Letters* **2014**, *14*, 1643**.**
[43]  P. Parkinson, H.J. Joyce, Q. Gao, H.H. Tan, X. Zhang, J. Zou, C. Jagadish, L.M. Herz, M.B. Johnston, *Nano Letters* **2009**, *9*, 3349.
[44]  J. Wallentin, M. Ek, L.R. Wallenberg, L. Samuelson, M.T. Borgström, *Nano Letters* **2011**, *12*, 151.
[45]  Y. Zhang, J. Wu, M. Aagesen, J. Holm, S. Hatch, M. Tang, S. Huo, H. Liu, *Nano Letters* **2014**, *14*, 4542-.
[46]  J. Wallentin, M. Ek, L.R. Wallenberg, L. Samuelson, K. Deppert, M.T. Borgström, *Nano Letters* **2010**, *10*, 4807.

[47]   R. Algra, M. Verheijen, M. Borgstrom, L. F. Feiner, G. Immink, W. van Enckevort, E. Vlieg, E. Bakkers, *Nature* **2008**, *456,* 369.
[48]   J. Wallentin, K. Mergenthaler, M. Ek, L.R. Wallenberg, L. Samuelson, K. Deppert, M. E. Pistol, M.T. Borgstrom, *Nano Letters* **2011**, *11*, 2286.
[49]   G. Stringfellow, W. Koschel, F. Briones, J. Gladstone, G. Patterson, *Applied Physics Letters* **1981**, *39*, 581.
[50]   K. Kudo, Y. Makita, I. Takayasu, T. Nomura, T. Kobayashi, T. Izumi, T. Matsumori, *Journal of Applied Physics* **1986**, *59*, 888.
[51]   Y. Fu, M. Willander, G. Chen, Y. Ji, W. Lu, *Applied Physics A: Materials Science & Processing* **200**4, *79*, 619.
[52]   C.K. Yong, K. Noori, Q. Gao, H.J. Joyce, H.H. Tan, C. Jagadish, F. Giustino, M.B. Johnston, L.M. Herz, *Nano Letters* **2012**, *12*, 6293.
[53]   O. Demichel, M. Heiss, J. Bleuse, H. Mariette, A. Fontcuberta i Morral, *Applied Physics Letters* **2010**, *97*, 201907.
[54]   M. Van Weert, O. Wunnicke, A. Roest, T. Eijkemans, A. Yu Silov, J. Haverkort, G. T Hooft, E. Bakkers, *Applied Physics Letters* **2006**, *88*, 043109.
[55]   Y. Hu, Y. Chang, P. Fei, R.L. Snyder, Z.L. Wang, *ACS Nano* **2010**, *4*, 1234.
[56]   X. Xie, S. Y. Kwok, Z. Lu, Y. Liu, Y. Cao, L. Luo, J.A. Zapien, I. Bello, C. S. Lee, S. T. Lee, *Nanoscale* **2012**, *4*, 2914.
[57]   H. Fang, W. Hu, P. Wang, N. Guo, W. Luo, D. Zheng, F. Gong, M. Luo, H. Tian, X. Zhang, *Nano Letters* **2016**, *16*, 6416.
[58]   J. Ohsawa, K. Saigoh, S. Yamaguchi, M. Migitaka, *Japanese Journal of Applied Physics* **1998**, *37*, 4758.
[59]   W. Spicer, Z. Liliental Weber, E. Weber, N. Newman, T. Kendelewicz, R. Cao, C. McCants, P. Mahowald, K. Miyano, I. Lindau, J*ournal of Vacuum Science & Technology B: Microelectronics Processing and Phenomena* **1988**, *6*, 1245.
[60]   B. Mukherjee, E.S. Tok, C.H. Sow, *Journal of Applied Physics* **2013**, *114*, 134302.
[61]   X. Wang, Y. Zhang, X. Chen, M. He, C. Liu, Y. Yin, X. Zou, S. Li, *Nanoscale* **2014**, *6*, 12009.
[62]   J. Suehiro, H. Imakiire, S.-i. Hidaka, W. Ding, G. Zhou, K. Imasaka, M. Hara, *Sensors and Actuators B: Chemical* **2006**, *114*, 943.

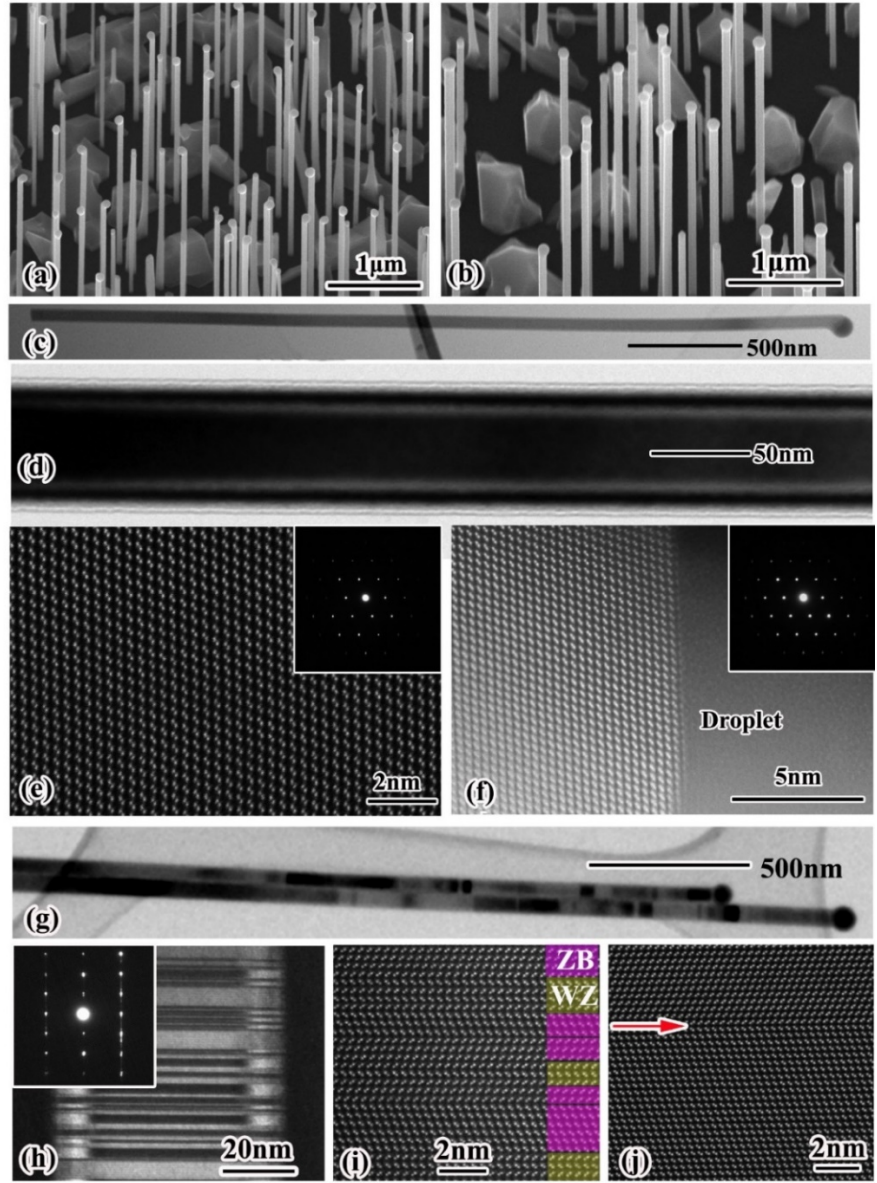

**Figure 1:** 30 degree tilted SEM image of self-catalyzed (a) p-doped and (b) un-doped GaAs NWs. (c)-(f) TEM images of p-NWs. (c) Low magnification TEM image of a whole p-NW that does not show any defect-related contrast difference. (d) Higher magnification TEM image shows only a segment of NW. Atomic resolution ADF-STEM image of (e) NW body and (f) tip of the NW. The insets in (e) and (f) are the electron diffraction patterns which confirm the pure-ZB crystal structure. (g)-(j) TEM images of un-doped NWs. (g) Low magnification TEM image of two NWs that show high-density defect-related contrast difference. (h) Higher magnification TEM image of a segment that show high-density of defects. The inset is the electron diffraction pattern which confirms the poly-type crystal structure. Atomic resolution ADF-STEM image of (i) mixture of zinc blende and wurtzite crystal structure and (j) single twin planes.

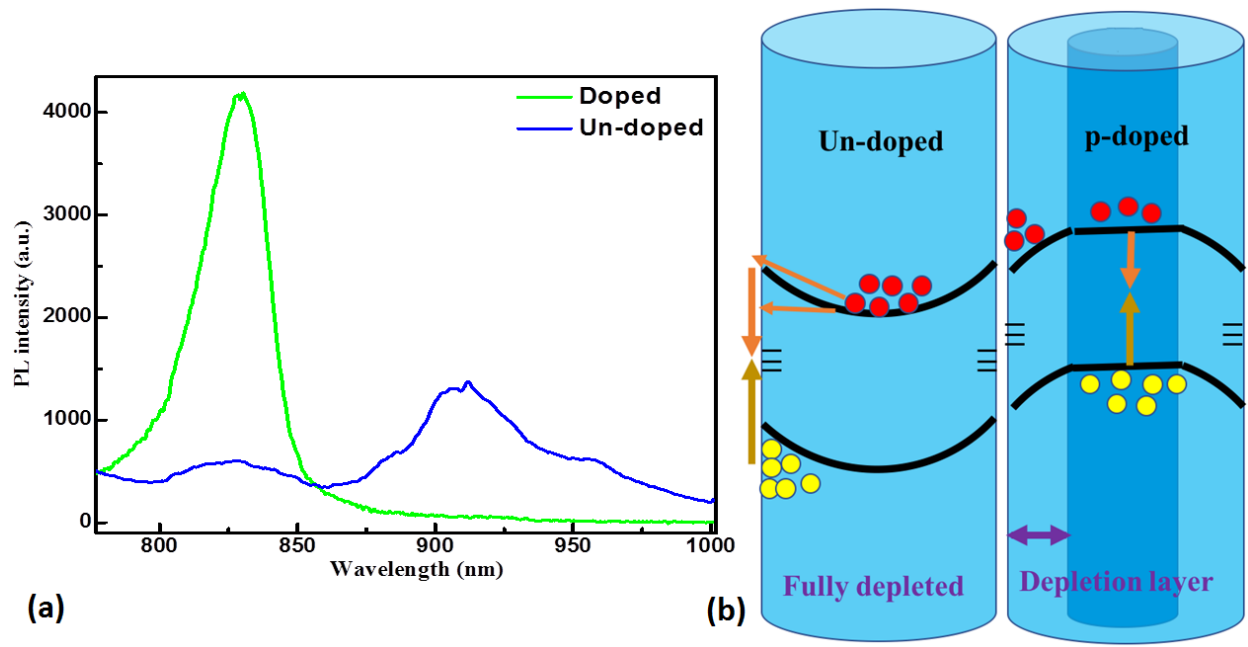

**Figure 2:** (a) Single NW PL spectra of doped and un-doped GaAs NWs at 20 K. (b) Schematic depicting band-bending effects across the nanowire cross section for (left) un-doped and (right) p-doped GaAs NWs.

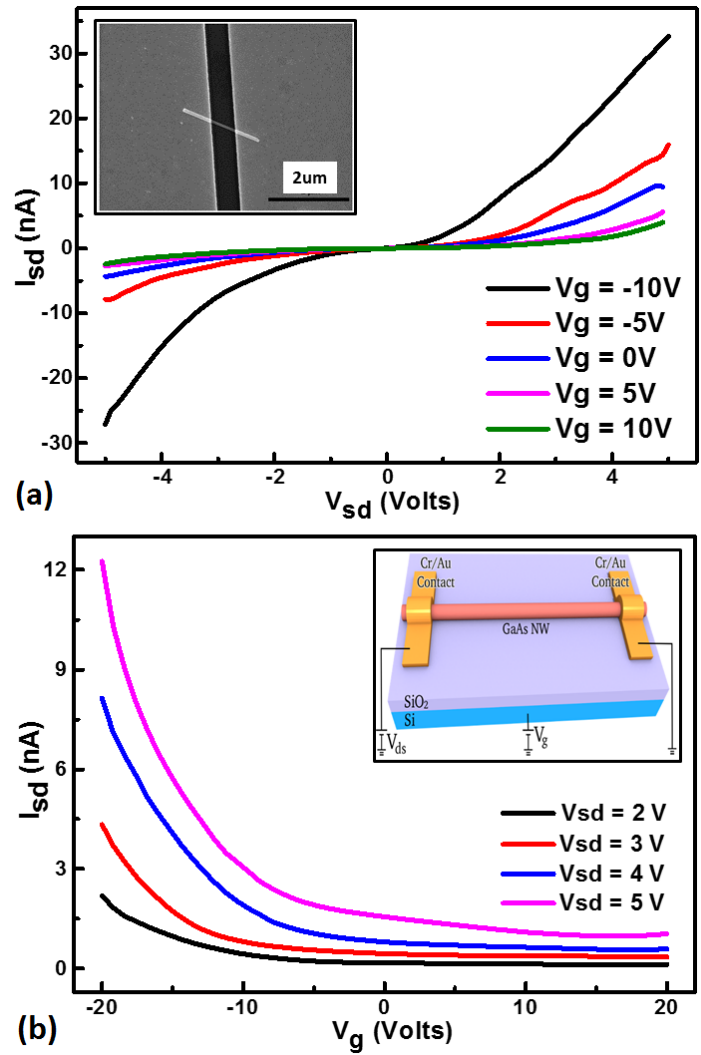

**Figure 3:** (a) Output characteristics of GaAs FET for different gate bias at room temperature. Inset shoes SEM image of GaAs NW Photodetector. (b) The transfer characteristics of GaAs NW FET for different drain voltages under forward sweep at room temperature. Inset shows the schematic diagram of GaAs single NW field effect transistor, doped Si is used as back gate electrode.

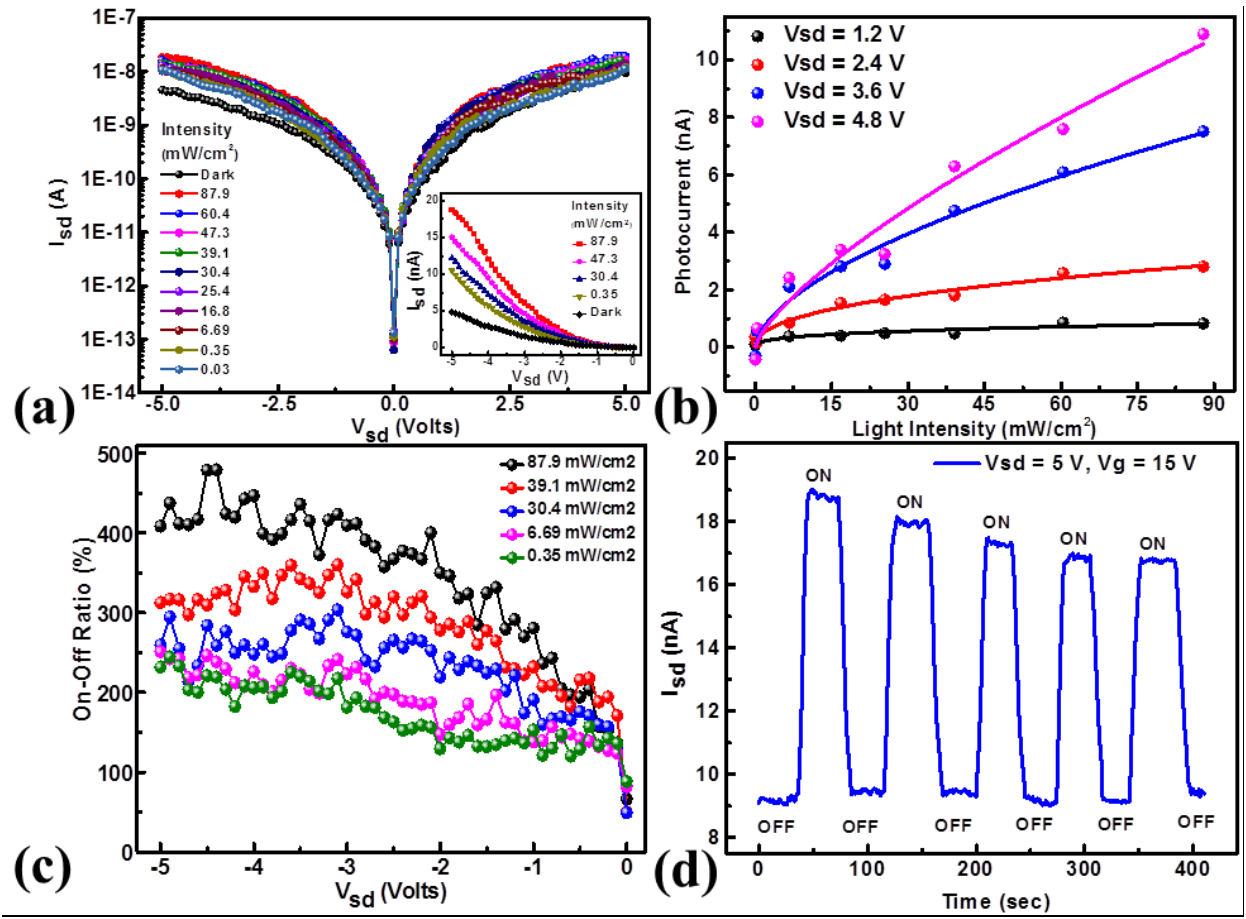

**Figure 4:** (a) The output characteristics curves of GaAs single NW photodetector under light illumination wavelength of 632.8 nm for different intensities, at gate bias of zero volt; inset shows the current on linear scale. (b) The photocurrent variation at zero gate bias with light intensity for various bias voltages following the power law. (c) The ratio of photocurrent to dark current with bias voltage for several light intensities at zero gate bias. (d) The switiching on and off response of GaAs NW PD for laser intensity of 87.9 mW/cm² revealing stablilty and repeatability of photodetector.

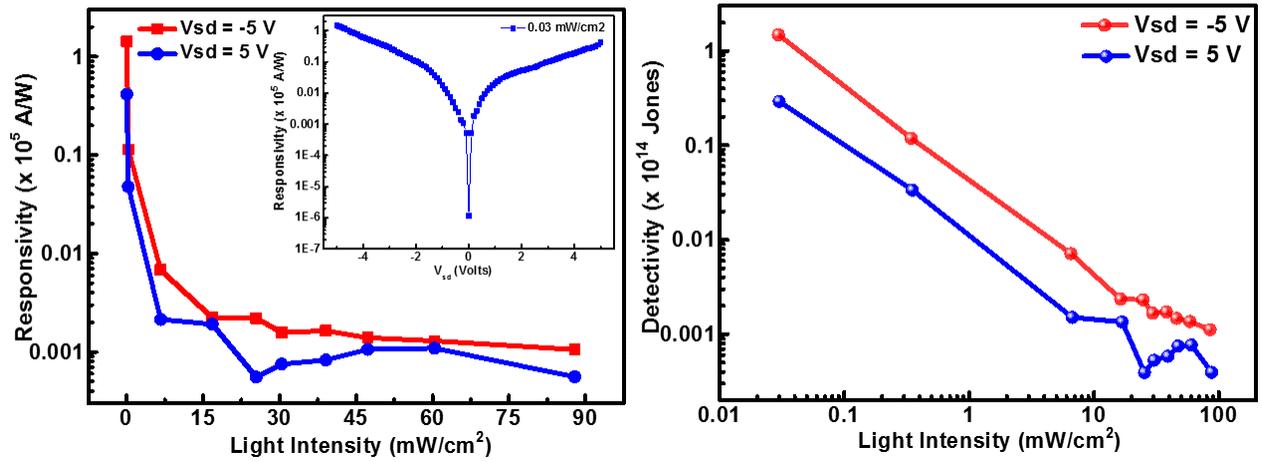

**Figure 5:** (a) Variation of spectral responsivity of GaAs NW PD with light intensity under different bias voltage; Inset shows the variation of responsivity of PD with applied bias under constant light intensity of 0.03 mW/cm$^2$. (b) The dependence of specific detectivity of GaAs NW PD on light intensity on double logarithmic scale.

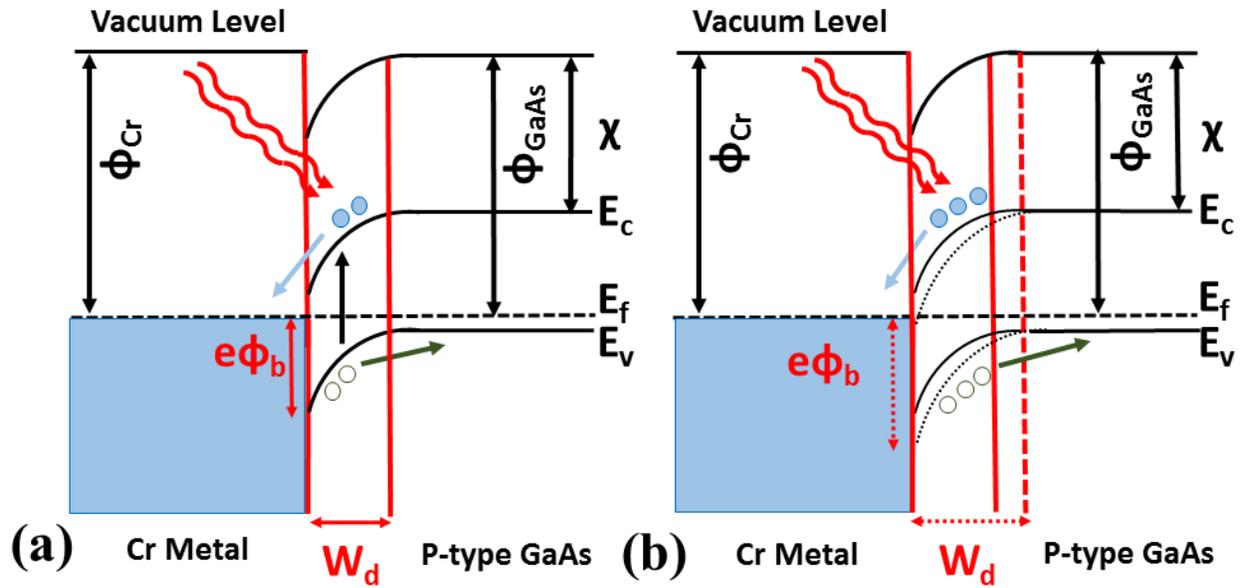

**Figure 6:** (a) Band diagram of metal-semiconductor schottky junction at thermal equilibrium. (b) Schottky junction under reverse biased with enhance width of depletion region, increasing photocurrent.